\documentclass[twocolumn,longbibliography,superscriptaddress]{revtex4-2} 

\usepackage{graphicx,color,appendix,physics,wasysym,xcolor,xparse}
\usepackage{setspace,appendix} 
\usepackage{amsmath,amssymb,amsfonts,mathtools}
\usepackage{bm,bbm}
\usepackage{euscript}
\usepackage{braket}
\usepackage{hyperref}
\hypersetup{colorlinks,linkcolor={red},citecolor={blue},urlcolor={blue}}

\newcommand{\bla}{\color{black}}

\newtheorem{definition}{Definition}

\usepackage{tikz}
 
\usepackage{verbatim,multirow}

\usepackage[utf8]{inputenc}
\usepackage[T1]{fontenc}
\usepackage{lmodern}

\newcommand{\mat}[2]{\left( \begin{array}{#1} #2  \end{array}\right)}

\newcommand{\ii}{\mathrm{i}}
\newcommand{\ee}{\mathrm{e}}

\begin{document}
	\title{\textbf{Detecting and quantifying non-Markovianity via quantum  direct cause}}
 \author{Shrikant Utagi}
 \email{shrikant.phys@gmail.com}
 \affiliation{Department of Physics, Indian Institute of Technology Madras, Chennai 600036, India.}
  \author{Prateek Chawla}  
\affiliation{Department of Applied Physics, Aalto University, FI-00076 Aalto, Finland.}
	
\begin{abstract}
	We study the efficacy of the two recently introduced witnesses of non-Markovianity, namely that based on temporal correlations in pseudo-density matrix and temporal steering correlations in detecting information backflow. We show, through specific counterexamples taken from existing literature, that they can witness a process to be non-Markovian where trace distance and entropic distinguishability measures may fail. We further show that, since the pseudo-density matrix is directly related to the Choi matrix of a channel via the partial transpose, it can be generalized to quantify the total quantum memory in any indivisible process. Moreover, we make an interesting observation that temporal steerable correlations-based measure may not capture eternal non-Markovianity hence may not be proportional to Choi-matrix-based methods, while pseudo-density matrix-based measures introduced in this work faithfully capture eternal non-Markovianity. Our work highlights important distinction between weak and strong forms of quantum direct cause in quantum mechanics when applied to open system dynamics.
\end{abstract}
\maketitle	

\section{Introduction}
No quantum system is perfectly isolated from its surroundings and inevitably undergoes the process of decoherence and dissipation \cite{breuer2002theory}, and such a system is called \textit{open} quantum system, whose dynamics is governed by quantum master equations or equivalently represented by a quantum channel -- a completely positive trace preservation (CPTP) map that takes density operators to density operators \cite{nielsen_chuang_2010}. 

Non-Markovian open system dynamics has been of interest for many years from both physics and quantum information theory point of view \cite{breuer2002theory,RHP14,breuer2016colloquium,vega2017dynamics,li2018concepts,shrikant2023quantum}.  Two broad notions of non-Markovianity are based on divisibility  \cite{RHP10}, which relies on the concept of entanglement, and on distinguishability \cite{breuer2009measure}, which relies on the concept of distance, for instance trace distance, between two states under a channel. Despite the fact that there are a plethora of non-Markovianity witnesses and measures, they are highly context-dependent. \cite{chruscinski2011divisiblity,chruscinski2014degree,chruscinski2017detecting,chruscinski2018divisibility}. However, for certain class of dynamical maps, namely the image non-increasing maps, they are indeed equivalent \cite{chakraborty2019information}. Interestingly, it has been established that the trace distance as an identifier of non-Markovianity essentially fails in three cases: (i) for eternally non-Markovian channel \cite{hall2014canonical} (ii) when non-unital part of the channel is solely responsible for the information backflow \cite{liu2013nonunital}, and (iii) when the dynamical map is trace-decreasing \cite{filippov2021trace}. Thus, a distance measure to witness P-indivisibility has to be chosen with some care. Not only the distance measures, even the system-ancilla entanglement would not witness CP-indivisibility when post-selection is incorporated for trace-decreasing maps \cite{filippov2021trace}. 

Quantum correlations, such as entanglement and steering, are those that cannot be realized by any local realistic theory, hence they are proven to be resources for quantum information and quantum computation tasks \cite{nielsen_chuang_2010,horodecki2009quantum,uola2020quantum}. Spatial correlations are generally known to be the correlations with a common cause \cite{feix2017quantum,gachechiladze2020quantifying}. Recent developments reveal that quantum mechanics allows for quantum correlations with direct cause dubbed temporal quantum correlations that arise out of measurements of observables across time. It has been shown \cite{ku2018hierarchy} that temporal quantum correlations form a hierarchy, namely temporal non-separability \cite{fitzsimons2015quantum,pisarczyk2019causal}, temporal steering \cite{chen2014temporal,Karthik15joint}, and temporal nonlocality \cite{leggett-garg1985inequallity,emary2013leggett}, and that temporally non-separable correlations in pseudo-density matrix (PDM) are a form of correlations with stronger quantum direct cause while temporal steerable correlations can be interpreted as those with weaker form of quantum direct cause. In fact, studies have been done to understand the fundamental differences between representation of temporally and spatially correlated states \cite{horsman2017can}, which motivates one to characterize their dynamics in physical settings in that we study their dynamics in open quantum systems. To mention in the passing, note that correlations present in PDM have relevance in physical situations such as ciruit QED \cite{lin2020quantum}, in certifying quantum channels \cite{shrotriya2024certifying} and in determining `temoprality' of a process \cite{song2024causal}, and temporal steering finds its application in quantum cryptography \cite{bartkiewicz2016temporal}. The full hierarchy of temporal correlations have been observed in superconducting circuits by manipulating the dynamics of the qubit \cite{weng2025observation}.

Recently, temporal steerable weight and causality measure defined in the PDM framework were used to define measures of non-Markovianity \cite{chen2016quantifying,shrikant2021causal}. One may wonder if these quantities are equivalent indicators of information backflow, since in the case of spatial correlations two different quantifiers of entanglement may be in-equivalent in detecting correlation backflow \cite{neto2016inequivalence}, while there may exist a measure that detects ``almost'' all non-Markovian dynamics \cite{santis2019correlation}. In this work, we address the question of the ability of temporal steering and temporal non-separable correlations to detect non-Markovianity of a generic quantum channel and establish their place in the divisibility hierarchy \cite{chruscinski2014degree,chruscinski2017detecting}.

Before proceeding further, it is important to mention a few restrictive assumptions that we employ in defining a dynamical map in this work. It is assumed that the two-time map arises when the environment degrees are traced out the system-environment (S-E) state is a product. Else, the map may be not-completely positive \cite{pechukas1994reduced,alicki1995comment,pechukas1995pechukas}. Although we will talk about \textit{intermediate} map being not CP, and even negative, the full map is always CPTP. It is just that the full channel can no more be decomposed into physically well-defined maps. This kind of picture may be thought of a black-box approach to understanding non-Markovian dynamics, where the full map is legit but what happens inside the black-box remains inaccessible. Furthermore, we are restricting to two-time correlations treatment of a dynamical process under this assumption, while there exist recent proposals such as process tensor \cite{costa2016quantum,pollock2018non,pollock2018operational,milz2021quantum}, conditional past-future correlations \cite{budini2018quantum}, and process matrix \cite{milz2018entanglement,giarmatzi2021witnessing} for characterizing non-Markovian dynamics taking into account the multi-time correlations. Nonetheless, it would be an interesting exercise to map process tensor \cite{pollock2018non,milz2021quantum} to a multi-time multi-qubit PDM evolved under a given quantum channel \cite{liu2023quantum}. It is also important to mention that constructing PDM experimentally may be resource intensive. However, it is possible to map the negativity of PDM to purity of a two-qubit state thereby allowing for an efficient detection of temporal correlations, see e.g,

This paper is structured as follows. In Sec.~(\ref{sec:div}), we discuss the well-known notions of CP-indivisibility and P-indivisibility (or, information backflow). In Sec.~(\ref{sec:pdm}) and (\ref{sec:steering}), we introduce temporal quantum correlations in PDM framework and temporal steering, respectively. In particular, in this work PDM framework is further generalized to continuous-time limit and a measure of CP-indivisibility is proposed. In Sec.~(\ref{sec:entropic}), we introduce the recently proposed entropic distance measure for non-Markovianity. We take up few examples that help us distinguish the performance of these measures. In Sec.~(\ref{sec:nonunitalNM}) we introduce the notion of non-unital non-Markovianity and define a purely non-non-unital non-Markovian channel and discuss the results. Then we briefly introduce the phase-covariant channel in Sec.~(\ref{sec:phase-cov}), that represents a most generic type of dynamics and discuss the results. Finally, we conclude in Sec.~(\ref{sec:discon}).

\subsection{Divisibility \label{sec:div}}
Let $\rho \in \mathcal{B}(\mathcal{H})$, where $\mathcal{B}(\mathcal{H})$ is the bounded operator space and $\Lambda: \mathcal{B}(\mathcal{H}) \rightarrow \mathcal{B}(\mathcal{H})$ be a quantum channel, with the operator-sum form $\Lambda[\rho] = \sum_j K_j \rho K_j^\dagger$, where $K_j$ are called the Kraus operators with $\sum_j K_j^\dagger K_j = \openone$. The map $\Lambda(t)$ is positive when it outputs a positive semidefinite state \textit{i.e.}, $\Lambda(t)[\rho] \ge 0$, for all $\rho \in \mathcal{B}(\mathcal{H})$. The map $\Lambda(t)$ is then said to be completely positive (CP) if the matrix 
\begin{align}
    \chi(t) = (I \otimes \Lambda(t))[\ketbra{\Psi}{\Psi}], 
    \label{eq:choi-state}
\end{align}
called the Choi matrix, is positive semidefinite \textit{i.e.}, $\chi(t) \ge 0$, where $I$ is the identity map and $\ket{\Psi} = \ket{00} + \ket{11} \in \mathcal{H} \otimes \mathcal{H}$ is an unnormalized maximally entangled state. 

The dynamical maps have some interesting properties. First, when the map $\Lambda(t+s) = e^{(t+s) \mathcal{L}} = e^{t \mathcal{L}}e^{s \mathcal{L}} = \Lambda(t) \Lambda(s)$ for all $t \ge s \ge 0$, then $\Lambda$ belongs to a family of quantum dynamical semigroups. Interestingly, semigroup dynamics may be categorized as an L-divisible process \cite{davalos2019divisibility}. However, the time-local generator $\mathcal{L}$ can be time-dependent with time-dependent decay rates as coefficients. In that case, assuming that the initial system-environment state is a product \cite{li2018concepts}, with rotating-wave approximation \cite{breuer2002theory} the dynamics is given by the generalized Gorini-Kossakowski-Sudarshan-Lindblad (GKSL)-like master equation of the form \cite{lindblad1975completely,sudarshan1961stochastic,hall2014canonical},
\begin{align}
  \frac{d \rho(t)}{dt} &=  \mathcal{L}(t)[\rho(t)]  \nonumber \\ &= \sum_j \gamma_j(t) \left(L_j(t) \rho(t) L^\dagger _j(t) - \frac{1}{2} \{L^\dagger _j(t)L_j(t), \rho(t) \}\right),
\end{align}
where $L_j(t)$ are called the time-dependent Lindblad operators. If the decay rates $\gamma_j(t) \ge 0$ for all $t$, then the following divisibility property holds \cite{RHP10}
\begin{align}
    \Lambda(t+\epsilon,0) = V(t+\epsilon,t)\Lambda(t,0), 
    \label{eq:CP-div}
\end{align}
for all $ t+\epsilon \ge  t \ge 0$. Assuming that some form of inverse $\Lambda^{-1}$ exists \footnote{In case when such an inverse does not exist, one may resort to Moore-Penrose pseudo-inverse \cite{RHP14}.}, one can define the intermediate map 
\begin{align}
    V(t+\epsilon,t) = \Lambda(t+\epsilon,0) \Lambda^{-1}(t,0).
    \label{eq:inverse}
\end{align}
Then the following definitions hold.
\begin{definition}
    A process is said to be CP-divisible if the intermediate map $V(t+\epsilon,t)$ is CP.
    \label{def:CP-div}
\end{definition}

\begin{definition}
    A process is said to be P-divisible if the intermediate map $V(t+\epsilon,t)$ is positive but may or may not be CP.
    \label{def:P-div}
\end{definition}
Note that a CP-divisible process is trivially P-divisible, but not the other way around.
\subsection{Distinguishability}
Given a distinguishability measure $ \mathcal{D}$, the Definition~(\ref{def:P-div}) is equivalent to
    \begin{align}
	\frac{d}{dt}\mathcal{D}(\Lambda(t)[\rho_1],\Lambda(t)[\rho_2]) \le 0
	\label{eq:Pindiv}
\end{align}
for any pair of $\rho_1,\rho_2 \in \mathcal{B}(\mathcal{H})$, where $\Lambda(t)$ is a quantum channel. Trace distance, given by $\mathcal{D}(\rho_1,\rho_2) = \frac{1}{2}\Vert \rho_1 - \rho_2 \Vert_1$, where $\Vert A \Vert_1 = {\rm Tr}\sqrt{A^\dagger A}$ is the trace norm, was proposed as a candidate in Eq.~(\ref{eq:Pindiv}) for the measure of non-Markovianity by Breuer-Laine-Piilo \cite{breuer2009measure}, while Hellinger distance would also do \cite{laine2010measure}. The non-monotonic behavior of any distance measure would be an indication of information backflow from the environment to system and such processes are P-indivisible \cite{chruscinski2011divisiblity,chruscinski2014degree,chruscinski2017detecting,chruscinski2018divisibility,chruscinski2022dynamical}.

It is known in the literature that the non-monotonicity of neither trace distance nor the recently introduced entropic distance are necessary for detecting information backflow \cite{liu2013nonunital,makela2015bounds,chruscinski2017detecting,megier2021entropic,smirne2022holevo,settimo2022entropic}, since they can be insensitive to certain parameter regimes. This can be clearly seen through the matrix representation of the dynamical map as follows. Any $d$-dimensional dynamical map $\Lambda(t)$ may be given a matrix representation as
\begin{equation}
F(t)=\left(\begin{array}{c|c}
1 & {\bm 0}_{1\times(d^{2}-1)}\\
\hline \bm{\kappa} & M
\end{array}\right),
\label{eq:mapmatrix}
\end{equation}
where $\bm{\kappa} \in \mathbb{R}^{d^2 -1}$ and $M$ is a $d^2-1 \times d^2 -1$ real matrix. Given the state $\rho=\frac{1}{d} (\openone + \sum_{i=1}^{d^2-1} r_i G_i) $, where $G_i$ are traceless orthonormal basis given by the generalized Gell-Mann matrices with $G_0 = \frac{\openone}{\sqrt{d}}$ and the Hermitian $G_i$ where $i \in \{1,.....d^2 -1\}$, the map $M_{ij}(t) := {\rm Tr}(G_i \Lambda[G_j])$, which transforms a Bloch vector $\bm{x}$ as
\begin{align}
    \bm{x} \rightarrow \bm{x}^\prime(t) = M(t)\bm{x}(0) + \bm{\kappa}. \label{eq:blochmap}
\end{align}
The components of the shift vector $\bm{\kappa}$ are obtained as $\kappa_i = {\rm Tr}(G_i\Lambda[\openone])$. A map is unital if all $\kappa_i = 0$, i.e., $\Lambda[\openone] = \openone$. If $\bm{x_1}$ and $\bm{x_2}$ are two Bloch vectors corresponding to the states $\rho_1$ and $\rho_2$, then from Eq.~(\ref{eq:mapmatrix}) and (\ref{eq:blochmap}), the transformation $\Lambda(t)[\rho_1 - \rho_2]$ corresponds to $M(t)[\bm{x_1} - \bm{x_2}]$. Therefore, the trace distance fails to witness non-Markovianity originating solely from $\bm{\kappa}$. However, the full non-Markovianity, in the sense of P-indivisibility is encoded in the map $F(t)$. Interestingly, the divisibility property of $\Lambda(t)$ carried over as $F(t) = F(t,s)F(s)$ leads to a non-trivial relationship between unital and non-unital parts of the channel \cite{chruscinski2017detecting}:
\begin{align}
   M(t) = M(t,s) M(s) \;\, \text{and} \;\, \bm{\kappa}(t) = \bm{\kappa}(t,s) + M(t,s) \bm{\kappa}(s),
\label{eq:nontrivial}  
\end{align}
where $\bm{\kappa}(t,s)$ and $M(t,s)$ parameterize $F(t,s)$. 

It is known that non-unitality of a channel is necessary for increase of purity $\mathcal{P} := {\rm Tr}(\rho^2)$. For instance, it can be shown that when $\Lambda(t)[\openone/2] \ne \openone/2$, that is when $\bm{\kappa} = \bm{0}$, then ${\rm Tr}(\Lambda(t)[\rho])^2 \le {\rm Tr}(\rho^2)$ for all $t$, even when the map $\Lambda(t)$ is unital and P-indivisible. Therefore, the full P-indivisibility is captured in $F(t)$ and one would require a distance measure that is sensitive to non-unital part. For simplicity, we will have the following definition. 
\begin{definition}
    When a dynamical map is indivisible solely due to the non-unital part of the channel then we call it ``non-unital non-Markovianity'' and such a channel is called purely non-unital non-Markovian channel.
\end{definition}

In this work we will show that the temporal quantum correlations are faithful in detecting information backflow for any unital and non-unital channel.

\section{Temporal quantum correlations and non-Markovianity}
\subsection{Pseudo-density matrix \label{sec:pdm}}
PDM construction captures spatial and temporal correlations in an equal footing \cite{fitzsimons2015quantum,pisarczyk2019causal}. There are a number of other such frameworks that address the problem of unification of spatiotemporal correlations, \textit{viz.} superdensity operator \cite{cotler2018superdensity}, the spatiotemporal doubled density operator \cite{jia2023spatiotemporal}, to name a few others. The frameworks such as process matrix \cite{oreshkov2012quantum}, process tensor \cite{pollock2018non} that describe most general stochastic processes \cite{chiribella2009theoretical} and closely resemble the framework of quantum combs, also capture non-Markovian processes from spatiotemporal correlation perspective. It has been shown that process matrix can be mapped to PDM in a number of ways \cite{zhang2020quantum}. Interestingly, PDM has been proposed to be relevant to the so-called causal inference problem \cite{liu2023quantum,song2024causal}.

In this section, we briefly introduce the two-point PDM and a measure of non-Markovianity based on it, and then extend the PDM formalism further on the level of Lindblad-like master equation. A two-point qubit PDM can be written as 
\begin{align}
    R(t) = I \otimes \Lambda(t) \bigg[\bigg \{\rho \otimes \frac{\openone_2}{2}, \mathcal{S} \bigg \}\bigg],
    \label{eq:PDM}
\end{align}
where $\mathcal{S} := \frac{1}{2} \sum_{i=0}^{3} \sigma_i \otimes \sigma_i$, $\sigma_i$ are Pauli operators with $\sigma_0 = \openone_2$, $\{A,B\} = AB + BA$, $I$ is the identity map and $\rho$ is the input to the PDM under the quantum channel $\Lambda(t)$. Fitzsimons et. al. introduced \cite{fitzsimons2015quantum} a measure of ``temporal-ness'' of correlations, dubbed causality measure  as the negativity of PDM, given by $f = \Vert R \Vert_1 - 1$. Later, it was generalized to a log negativity of PDM as 
\begin{align}
    F(t) = \log \Vert R(t) \Vert_1
    \label{eq:cmmeasure}
\end{align}
which preserves additivity property, and is shown to be related to the quantum capacity of a given channel \cite{pisarczyk2019causal}, where $R(t)$ is given by Eq.~(\ref{eq:PDM}). We call $F$  the logarithmic causality measure (LCM) to distinguish it from $f$. $F$ is nonincreasing under CPTP map and a measure of non-Markovianity has been defined \cite{shrikant2021causal} using $F$, for a given channel $\Lambda$ and input state $\rho$, as the integral over positive slope of $F$
\begin{align}
   \mathcal{N}^{\rm LCM} :=  \int_{\frac{dF}{dt}>0}  dt \; \frac{dF(\rho,\Lambda,t)}{dt}.
\label{eq:cmeasure0}
\end{align}

In this work, we further extend the ability of PDM to quantify non-Markovianity of any CP-indivisible process as follows.
It is known that a causal PDM - a PDM with temporal correlations - is nothing but partial transposed Choi matrix of a channel \cite{horsman2017can,zhao2018geometry} with an arbitrary input state $\rho$. 
Using the relation in Eq. (\ref{eq:inverse}) we then have
\begin{align}
    R_{t+\epsilon,t} = I \otimes V(t+\epsilon,t) \bigg[\bigg \{\rho \otimes \frac{\openone_2}{2}, \mathcal{S} \bigg \}\bigg].
    \label{eq:intermediatePDM}
\end{align}
Now, we define the following measure of total non-Markovianity 
\begin{align}
 \mathcal{N}^{ \rm {\tiny causality}} = \int_{0}^{\infty}dt \; \mu (t); \quad 
      \mu (t) := \lim_{\epsilon \rightarrow 0^+} \frac{\Vert R_{t+\epsilon,t} \Vert_1 - 1}{\epsilon}.
      \label{eq:PDM-RHPmeasure} 
\end{align}

\begin{figure}
    \centering
    \includegraphics[width=0.8\linewidth]{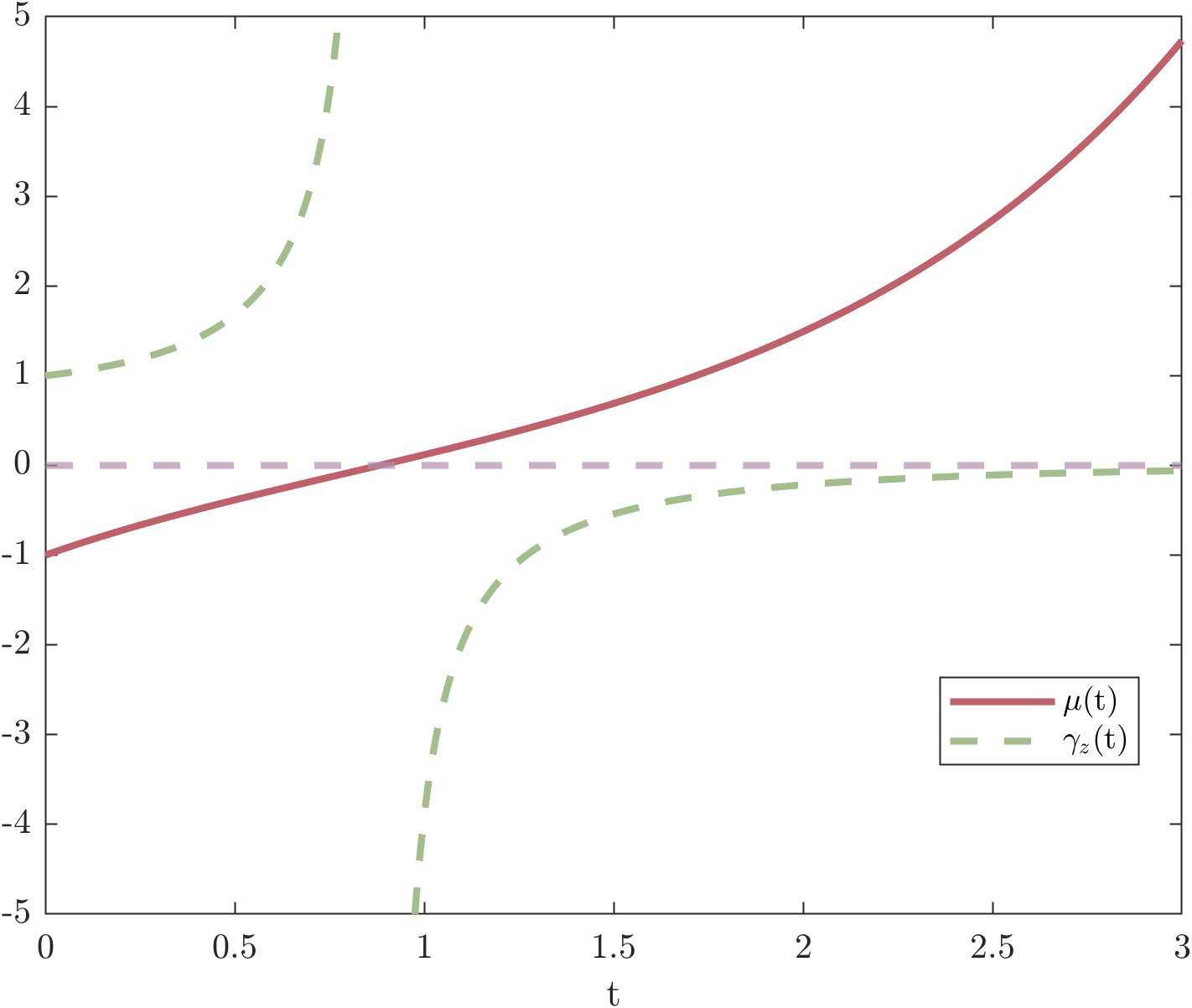}
    \caption{Figure showing the equivalence between negativity of $\gamma(t)$ and positivity of $\mu (t)$ for a dephasing channel with a decay rate given by  $(1 + \alpha e^{-t})/(2- 2\alpha \sinh t)$. The plot is shown for $\alpha = 5$.}
    \label{fig:newCMdecayrate}
\end{figure}

 For a simple dephasing dynamics, one can read out the decay rates by plugging $\Lambda(t+\epsilon,t) \rightarrow e^{\epsilon\mathcal{L}(t)}$ in Eq.~(\ref{eq:intermediatePDM}) in the limit $\epsilon \rightarrow 0$:
 \begin{align}
      \mu (t) = \underset{\epsilon \rightarrow 0^{+}} {\lim} \frac{ \bigg\Vert I_4 + \epsilon(\mathcal{L}(t) \otimes I_2 ) \bigg[\bigg\{\rho \otimes \frac{\openone_2}{2}, \mathcal{S} \bigg\}\bigg] \bigg\Vert_1 - 1}{\epsilon}
      \label{eq:causalmu(t)}
 \end{align}
 where $\mathcal{L}(t)[\rho] = \gamma_z(t) (\sigma_z \rho \sigma_z - \rho)$. Here, it suffices to take the Taylor expansion of $ e^{\epsilon \mathcal{L}(t)}$ up to the first order. Consider for example a decay rate of the form $\gamma_z(t) = \frac{1 + \alpha e^{-t}}{2(1- \alpha \sinh t)}$. One may find that whenever $\mu (t)$ is positive, $\gamma_z(t)$ is negative for the given interval. Figure~\ref{fig:newCMdecayrate} shows the negativity of decay rate being exactly equal to the positivity of $\mu (t)$ defined in Eq.~(\ref{eq:causalmu(t)}). 

 The witness for the CCM-based measure requires to be normalized appropriately, and we use the method detailed in Ref.~\cite{RHP14}, where the measure $\mathcal{N^{\text{CCM}}}$ is defined from the witness $\mu(t)$ as 

\begin{equation}
    \begin{split}
        \mathcal{N^{\text{CCM}}} &= \frac{\int_0^t \bar{\mu}(t)}{\int_0^t \chi(t)}, \\
        \text{where, } \bar{\mu}(t) &= \begin{cases}
            \tanh(\mu(t)) & \mu(t) > 0, \\
            0 &\text{otherwise.}
        \end{cases}\\ 
        \text{and } \chi(t) &= \begin{cases}
            0 & \bar{\mu(t)} = 0,\\
            1 & \text{otherwise.}
        \end{cases}
    \end{split}
    \label{eq:witness-norm-enm}
\end{equation}

We consider here that $0/0 = 0$. In practice, the condition is implemented by using $\chi(t)\rvert_{\bar{\mu(t)}= 0} = \epsilon $, where $\epsilon \rightarrow 0^+$.

 One must note that PDM is entirely constructed out of measurement statistics of product of Pauli operators over time. That is, for a given input state $\rho$, one performs a full quantum state tomography and then the state is evolved through a quantum channel, then one again performs the full state tomography. In fact, for $\rho = \frac{\openone}{2}$ in Eq.(\ref{eq:intermediatePDM}), $\mathcal{N}^{\rm {\tiny causality}}$ reduces to a well-known measure due to Rivas-Huelga-Plenio proposed in Ref. \cite{RHP10}, with a similar advantage that it does not require any sort of optimization. 

\subsection{Temporal steering \label{sec:steering}}
In the similar spirit to the spatial steering via a pre-shared entangled state, one can indeed steer a state in time using a maximally temporally correlated PDM. Here we describe temporal steering scenario and define the temporal steerable weight as an indicator of information backflow.

Alice possesses a state and performs a measurement on her state transforming an input state $\rho$ at $t=0$ into $\rho_{a|x} = (p(a|x))^{-1} \Pi_{a|x} \rho \Pi_{a|x}^\dagger$, where $\Pi_{a|x}$ is the conditional measurement and $p(a|x) := \text{Tr}(\Pi_{a|x} \rho \Pi_{a|x}^\dagger)$ is the probability that the outcome $a$ occurs when $x$ is measured. Then Alice sends her state down a quantum channel $\Lambda$. Bob has access to the temporal assemblage $\sigma_{a|x}(t) = \Lambda(t)[\sigma_{a|x}(0)]$ created by Alice via her measurements then evolved in time through $\Lambda(t)$. Now, Bob has no way of identifying whether the correlations are created due to an underlying cause, such as some hidden variables $\lambda$, or whether they are genuinely created due to Alice's measurements. Bob ultimately performs a quantum state tomography on his temporal assemblage and finds the state $\sigma_{a|x}(t)$. Let us call the assemblages that are genuinely created by Alice's measurement as temporally steerable, denoted by $\sigma^{\rm \small S}_{a|x}(t)$, and the ones created by an underlying hidden state model as unsteerable, defined as $\sigma^{US}_{a|x}(t) = \sum_\lambda P(\lambda) P(a|x, \lambda ) \sigma_\lambda$. If Bob is unable to write down his assemblage as $\sigma^{US}_{a|x}(t)$ then he is sure that Alice solely prepared his state and sent it via the channel. 

A measure of strength of temporal steering, called temporal steerable weight (TSW) was introduced \cite{chen2016quantifying} via the convex mixture 
\begin{align}
    \sigma_{a|x}(t) = w \sigma^{US}_{a|x}(t) + (1 - w) \sigma^{S}_{a|x}(t) \quad \forall a,x,
\end{align}
where TSW is defined by $W^{\rm TS} = 1 - w'$, with $w'$ is the maximum value of $w$. Here, $w'=1$ or $0$ corresponds to maximum and minimum steerability, respectively. One would find $w'$ via the dual semidefinite program:
\begin{align}
  \text{Find} \quad  w' &= \text{max} \, \text{Tr} \sum_\lambda  w \sigma_\lambda, \nonumber \\
    \text{subject \, to} \quad \bigg(&\sigma_{a|x}(t) - \sum_\lambda q_\lambda(a|x) w \sigma_\lambda \bigg) \ge 0 \quad \forall a,x \nonumber \\
    & w \sigma_\lambda \ge 0 \quad \quad \forall \lambda,
\end{align}
where $q_\lambda(a|x)$ are the extremal values of $P(a|x,\lambda)$, which are tabulated in Appendix. Under a noisy P-divisible quantum channel, TSW satisfies the inequality $W^{\rm TS}_\rho \ge W^{\rm TS}_{\Lambda[\rho]}$, where $\rho$ is some initial state. A measure of non-Markovianity or information backflow was defined in Ref.~\cite{chen2016quantifying} as
\begin{align}
\mathcal{N} := \int_{t_0}^{t_{\rm max}} \biggl| \frac{dW^{\rm TS}}{dt} \biggr|  dt + (W^{\rm TS}_{t_{\rm max}} -  W^{\rm TS}_{t_0} ).
\label{eq:TSmeasure2}
\end{align}
which can be normalized $N = \frac{\mathcal{N}}{1 + \mathcal{N}}$ to fit the range $\{0,1\}$. Here, maximum and no information backflow correspond to $N=1$ and $N=0$, respectively.

A MATLAB repository to implement quantum spatial steering can be found in Ref.~\cite{cavalcanti2016quantum}. In this work, one requires to replace the two-qubit measurements with measurements on a single qubit before and after the action of the channel $\Lambda(t)$. A python package to implement convex programming is available on \cite{cvxoptpython}.

\section{Entropic distinguishability measures \label{sec:entropic}}
Entropic distance measures of non-Markovianity would faithfully identify information backflow due to non-unital non-Markovianity \cite{megier2021entropic}, which was shown using a phase-covariant non-unital channel \cite{smirne2016ultimate}. Quantum relative entropy would be a candidate for the same, however due to its unboundedness, it cannot be used for witnessing of information backflow for arbitrary pairs of initial states since that would introduce singularities in the non-Markovianity measure. Therefore, a regularized version of it, called the telescopic relative entropy (TRE), can be a better candidate due to its boundedness \cite{audenaert2011telescopic}, which is given as
\begin{align}
    T_\nu (\rho_1 \Vert \rho_2 ) =  \frac{1}{\log \nu^{-1}} S (\rho_1, \nu \rho_1 + (1-\nu)\rho_2 ),
\end{align}
where the telescopic parameter $\nu$ represents the strength of mixing between the states $\rho_1$ and $\rho_2$. Here, $S(X,Y) := {\rm Tr}(X \log X - X \log Y)$ is the Umegaki quantum relative entropy between any two $X,Y \in \mathcal{B}(\mathcal{H})$. 

None of these entropic quantities induce a metric on the bounded operator space, hence are not proper distances. Nonetheless, the symmetrized TRE with $\nu = \frac{1}{2}$ is called the quantum Jensen-Shannon divergence (QJSD), whose square root is recently proven to be a distance \cite{megier2021entropic} given by 
\begin{align}
\mathcal{D}(\rho, \rho_2) = \frac{1}{\sqrt{2}} \bigg( T_{1/2} \big(\rho_1 \Vert \frac{\rho_1 +\rho_2}{2} \big) +  T_{1/2} \big(\rho_2 \Vert \frac{\rho_1+\rho_2}{2} \big) \bigg)^{\frac{1}{2}}.
\label{eq:qjsd}
\end{align}

\section{Examples}
\subsection{Purely non-unital non-Markovian channel \label{sec:nonunitalNM}}
Any qubit channel, up to a unitary transformation, is equivalently described by substituting $\bm{\kappa} = (0,0,\tau_3)^T$ and $M_{ij} = \{\lambda_1,\lambda_2,\lambda_3\}$ with $M_{i \ne j} = 0$ in Eq.~(\ref{eq:mapmatrix}). For a qubit, the Pauli operators $\sigma_i, \, i \in \{0,1,2,3\}$, with $\sigma_0 = \openone_2$, form the orthonormal Hilbert-Schmidt basis $\{G_i\}$ for $d=2$. A purely non-unital non-Markovian qubit generalized amplitude damping (GAD) channel was introduced in Ref.~\cite{liu2013nonunital} which in terms of the map $F(t)$ corresponds to the parameters $\lambda_1(t) = \lambda_2(t) = \sqrt{1 - \nu(t)}$, $\lambda_3(t) = 1- \nu(t)$ and $\tau_3(t) = (1-2p(t))\nu(t)$, where $p(t)= \sin^{2} t$ and $\nu(t) = 1 - e^{-t}$. In fact, for a GAD channel the trace distance evaluates to $\sqrt{1-\nu(t)}$, thus it clearly fails to witness non-Markovianity as the non-monotonicity originates from $p(t)$ rather than $\nu(t)$ \cite{liu2013nonunital}. However, certain entropic measures detect purely non-unital non-Markovianity \cite{megier2021entropic}.

\begin{figure}
    \centering
    \includegraphics[width=0.8\linewidth]{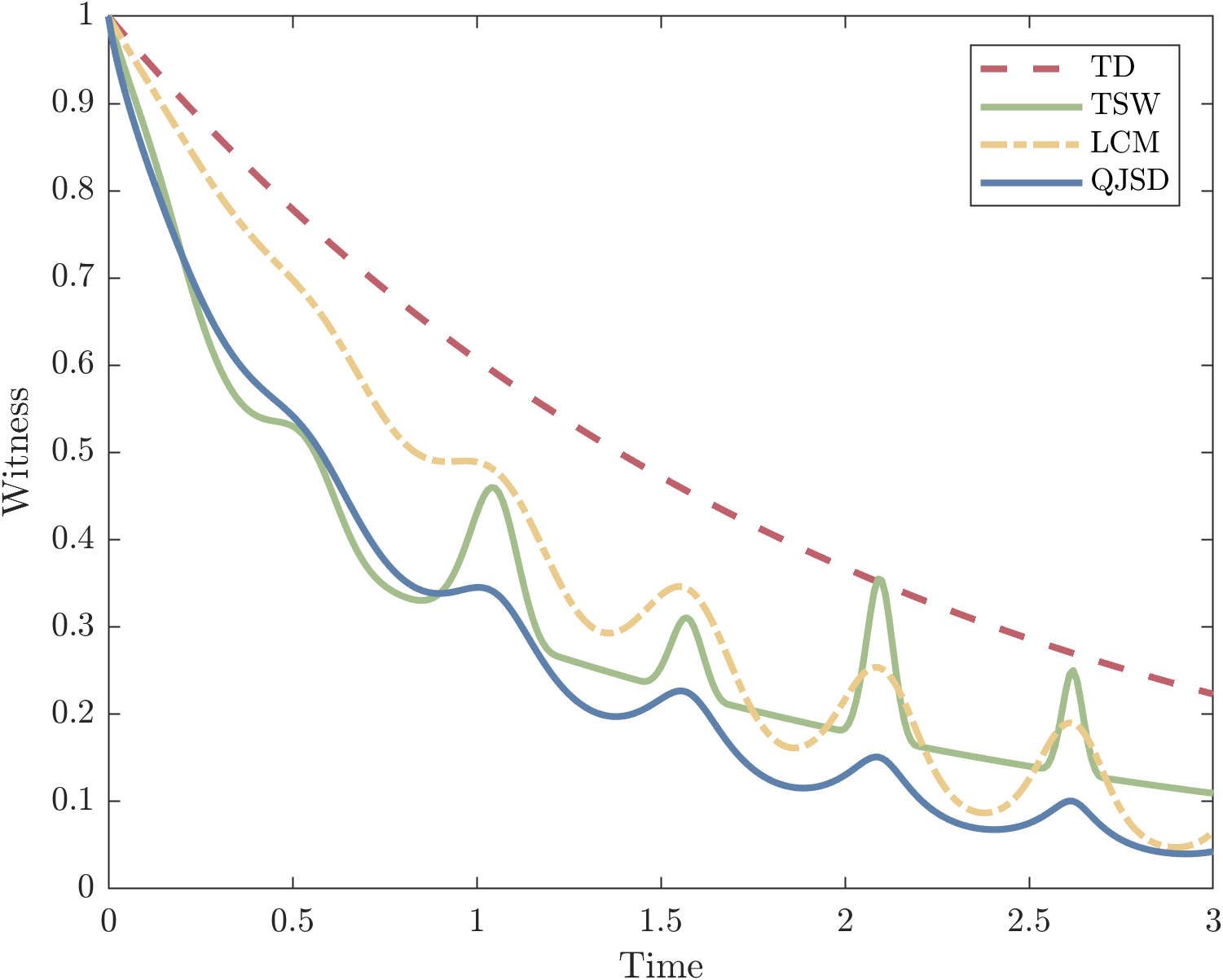}
    \caption{Various witnesses against a non-Markovian GAD channel. The witnesses for trace distance (dashed red), temporal steerable weight (solid green), quantum Jensen-Shannon divergence (solid blue), and the logarithmic causality (dotted yellow) measures for a non-Markovian GAD channel. The channel uses the parameter $p(t) = \sin^{2}(5t)$.}
    \label{fig:GADwitness}
\end{figure}

The Figure (\ref{fig:GADwitness}) shows the difference in the ability of various information-theoretic quantities in detecting purely non-unital non-Markovianity. \bla

\subsection{Phase-covariant channel \label{sec:phase-cov}}
 Here, we will use a phase-covariant channel  given in Refs.~\cite{filippov2020phase,smirne2016ultimate,settimo2022entropic}. And we will also use a variant of generalized amplitude damping channel given in \cite{neto2016inequivalence}.

One may define the canonical Kraus operators that satisfy $\Lambda[\rho]=\!\sum_i  K_i \rho K_i^{\dag}$ \cite{bengtsson2017geometry}:
\begin{align}
K_1 &= \sqrt{\tfrac{1-\eta_\parallel+\kappa}{2}} \mat{cc}{0 & 1 \\ 0 & 0} \nonumber\\
K_2 & = \sqrt{\tfrac{1-\eta_\parallel-\kappa}{2}} \mat{cc}{0 & 0 \\ 1 & 0} \nonumber\\
K_3 & = \sqrt{g_+} \mat{cc}{\cos\phi & 0 \\ 0 & \sin\phi \,\ee^{\ii \varphi}} \nonumber\\
K_4 & = \sqrt{g_-} \mat{cc}{-\sin\phi & 0 \\ 0 & \cos\phi \,\ee^{\ii \varphi}},
\label{eq:Kraus_ops}
\end{align}
where
\begin{align}
g_{\pm} & = \frac{1 + \eta_\parallel \pm \sqrt{\kappa^2 + 4\eta_\perp^2}}{2} \nonumber\\
\cot \phi & = \frac{\kappa + \sqrt{\kappa^2 + 4 \eta_\perp^2}}{2 \eta_\perp}. \label{eq:cot(theta)}
\end{align}

Eq.~(\ref{eq:Kraus_ops}) is a solution to master equation
\begin{align}
    \nonumber \frac{d}{dt}\rho(t) = &-\frac{i}{2}\left( \omega + h(t)\right) \left[ \sigma_z, \rho(t) \right] \\
    \nonumber &+ \gamma_+ \left( \sigma_+\rho(t)\sigma_- - \frac{1}{2} \left\{ \sigma_-\sigma_+, \rho(t) \right\} \right) \\
    \nonumber &+ \gamma_- \left( \sigma_-\rho(t)\sigma_+ - \frac{1}{2} \left\{ \sigma_+\sigma_-, \rho(t) \right\} \right) \\
    &+\gamma_z(t)\left( \sigma_z\rho(t)\sigma_z - \rho(t)\right)
\end{align}
where,
\begin{equation}
    \begin{split}
        \varphi &= \omega t + \theta \\
        h(t) &= \frac{d}{dt}\theta(t) \\
        \gamma_\pm(t) &= \pm\frac{1}{2} \left( \frac{d}{dt}\kappa(t) \mp \frac{1}{\eta_\parallel(t)}\frac{d}{dt}\eta_\parallel(t) \left( 1 \pm \kappa(t)\right) \right) \\
        \gamma_z(t) &= \frac{1}{4} \left( \frac{1}{\eta_\parallel(t)}\frac{d}{dt}\eta_\parallel(t) - 2\frac{1}{\eta_\perp(t)}\frac{d}{dt}\eta_\perp(t) \right)
    \end{split}
\end{equation}
\bla
Specifically, many standard qubit channels (see \cite{bengtsson2017geometry}) can be seen as special cases of the map \( \Lambda \). For example: pure dephasing is obtained when \( \eta_\parallel = 1 \), \( \kappa = 0 \), and \( \eta_\perp > 0 \); isotropic depolarization (i.e., white local noise) corresponds to \( \eta_\parallel = \eta_\perp > 0 \) and \( \kappa = 0 \); amplitude damping is described by \( 0 \leq \kappa \leq 1 \), with \( \eta_\parallel = 1 - \kappa \) and \( \eta_\perp = \sqrt{1 - \kappa} \). Also, the rotation angle is given by \( \varphi = \omega t + \theta \).

Now, for \textit{unital} phase-covariant channels -- those that preserve the identity state (\textit{i.e.}, \( \Lambda[\tfrac{1}{2}\openone] = \tfrac{1}{2}\openone \)) -- this occurs when there is no displacement, \textit{i.e.,} \( \kappa = 0 \). In such unital cases, the Kraus operators (see Eq.~\eqref{eq:Kraus_ops}) become notably simpler, especially when \( \phi = \pi/4 \).
\bla 

Now, we consider an example of a dynamical map $\Lambda$ of a counterexample that is non-Markovian according to trace distance but is Markovian according to the distance based on QJSD, with the following parameterization of the phase covariant channel.
\begin{gather}
  \label{eq:couterexample-eta}
  \begin{split}
    \eta_{\parallel,\perp}(\tau) =& e^{-\mu_1 \tau}\sigma(1-\tau) \\
    &+ e^{-\mu_1}e^{-\mu_2(\tau-1)}\sigma(\tau-1)\sigma(2-\tau) \\
    &+ e^{-\mu_1-\mu_2}\left[(3-\tau)+A_{\parallel,\perp}(\tau-2)\right]\sigma(\tau-2),
  \end{split}\\
  \label{eq:couterexample-kappa}
  \begin{split}
      \kappa_z(\tau) =& A_\kappa \tau \sigma(2-\tau) \\&+ 2 A_\kappa \left[(3-\tau)+A_{\parallel}(\tau-2)\right]\sigma(\tau-2),
  \end{split}
\end{gather}
where $\sigma$ is the sigmoid function
\[
  \sigma(\tau) = \frac1{1+e^{-\alpha \tau}},
\]
which is a smooth version of the Heaviside $\theta$ function and $\tau = t/T$ is a dimensionless time parameter, where $T$ is a reference time determining the duration of the different stages of dynamics.  Here, $A_{\perp}, A_{\parallel}, A_{\kappa}$ are real numbers.

\begin{figure}
    \centering
    \includegraphics[width=0.8\linewidth]{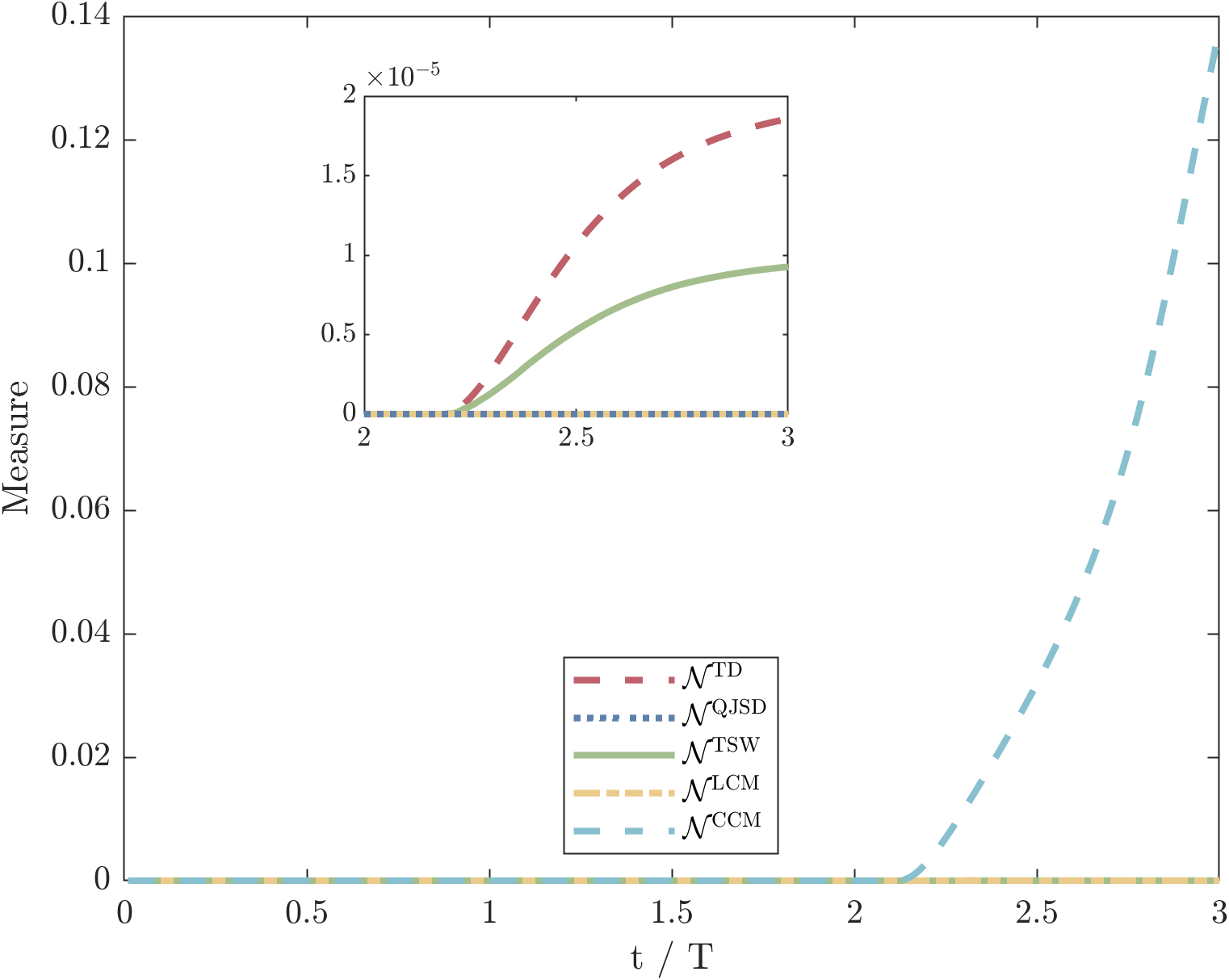}
    \caption{The non-Markovianity measures corresponding to trace distance, QJSD, temporal steerable weight, and both the logarithmic and continuous causality measures for a phase-covariant channel. The channel is characterized as in Eqs.~\eqref{eq:couterexample-eta} and \eqref{eq:couterexample-kappa}, with $A_\parallel = 0.01$, $A_\perp = 1.01$, $\mu_1 = 5$, $\mu_2 = 4$, and $\alpha = 5$.} 
    \label{fig:phasecovFig1}
\end{figure}

\begin{figure}
    \includegraphics[width=0.68\linewidth]{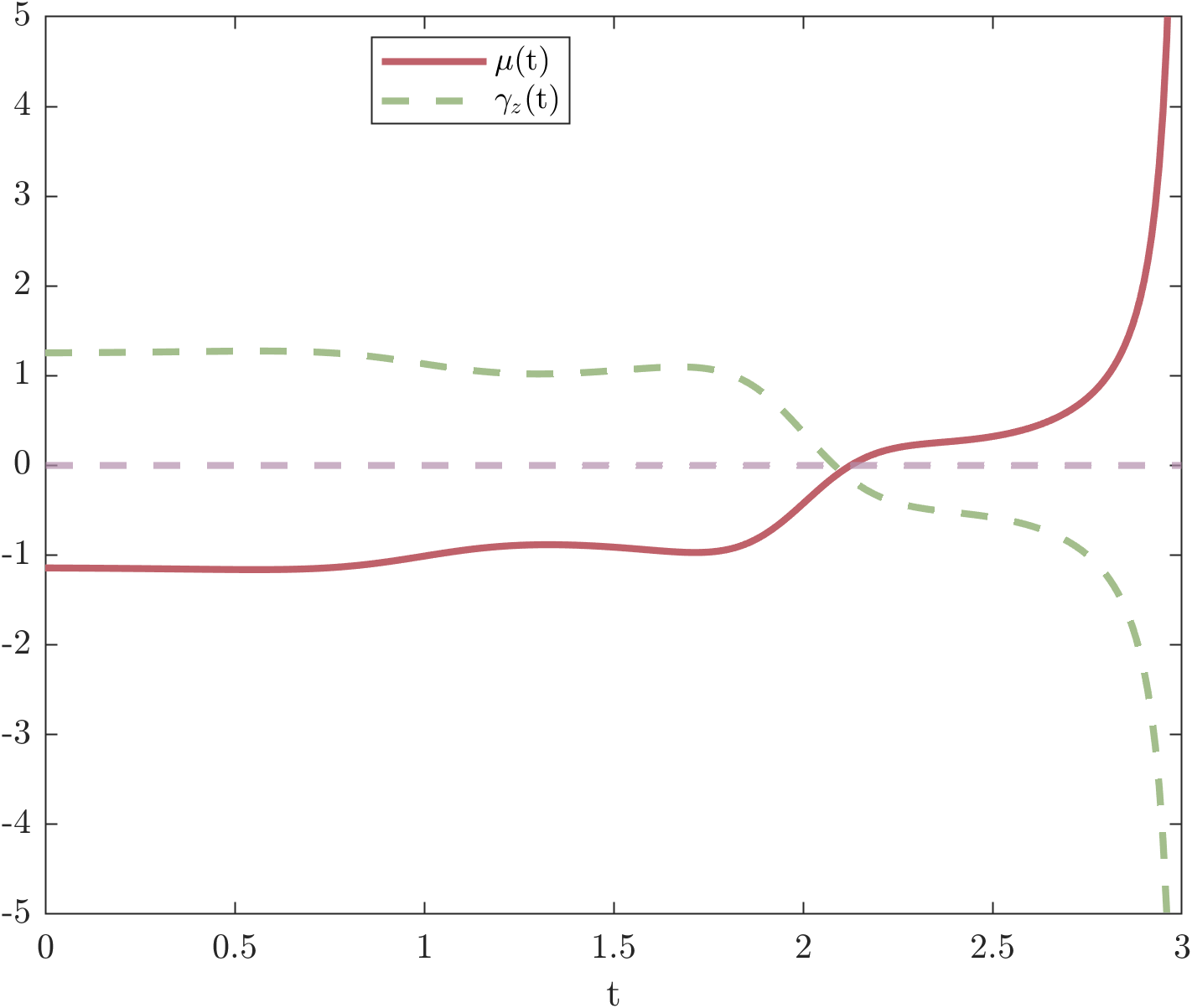}
    \caption{$\mu (t)$ under phase-covariant channel. The CCM-based witness $\mu(t)$ and the decay rate $\gamma_z(t)$ for the phase covariant channel. The specific characterization of the channel is the same as that of Fig.~\ref{fig:phasecovFig1}. }
    \label{fig:phasecov-g-witness}
\end{figure}

We benchmark the performance of various quantities considered in this paper against this channel. The results are given in Figure (\ref{fig:phasecovFig1}). It is seen that the trace distance and the temporal steerable weight measures are able to detect the non-Markovian character of the channel, as well as the continuous form of the causality measure. The logarithmic form is not able to detect the non-Markovianity of the channel, which implies it is a stronger measure than the continuous case. In Figure (\ref{fig:phasecov-g-witness}) it is shown that CCM-based measure quantifies negativity of decay rate for the phase-covariant channel.

\subsection{Eternally non-Markovian unital channel}
One talks about eternal non-Markovianity when there is at least one decay rate that remains negative at all times. However, there may exist a channel with two decay rates which periodically take on negative values such that the effective dynamics of a qubit becomes eternally non-Markovian. For instance, a superconducting transmon qubit may undergo such a process under some physical settings as reported in Ref.~\cite{gulacsi2023signatures}. Here, we consider a toy channel introduced in Ref.~\cite{hall2014canonical} which is eternally non-Markovian in the sense that it consists of three decay rates out of which one remains negative at all times $t > 0$. Consider the following generator
\begin{align}
&\mathcal{L}(t)[\rho] = \sum_{i=1}^{3} \gamma_i(t) (\sigma_i \rho \sigma_i - \rho ) \nonumber \\
\text{with} \; &\gamma_1=\gamma_2=\frac{c}{2} \; \text{and} \; \gamma_3(t) = - \frac{c}{2}\tanh t,
\label{eq:ENM-channel}
\end{align}
where $\sigma_i$ are the Pauli operators, and $\gamma_i(t)$ are the canonical decay rates, where $c$ is some real constant. The solution of the equation $\frac{d}{dt}\Lambda(t) = \mathcal{L}(t)\Lambda(t)$ leads to the corresponding dynamical map $\Lambda(t)[\rho] = \mathcal{T} \exp \{ \int_{t_0}^{t} ds \mathcal{L}(s)\}[\rho] = \sum_i E_i(t) \rho E_i^\dagger(t)$ where $E_i(t)$ are the Kraus operators given by \cite{wudarski2016markovian} \bla
\begin{equation}
E_1 = \sqrt{2 k(t)}I,~~ E_2 = \sqrt{k(t)}\sigma_1; ~~E_3 = \sqrt{k(t)}\sigma_2,
\label{eq:ENM}
\end{equation} 
where $k(t) := \frac{1}{4}(1-e^{-c t})$. One can straightforwardly go from time-local generator to the dynamical map and vise-versa via the so-called $H$-matrix method, for Pauli channels \cite{chruscinski2013non,chruscinski2015non}. Clearly, this channel is eternally CP-indivisible as $\gamma_3(t)$ remains negative for all $t> 0$. However, it is P-divisible since the intermediate map remains positive for all $t$.

\begin{figure}
    \centering
    \includegraphics[width=0.8\linewidth]{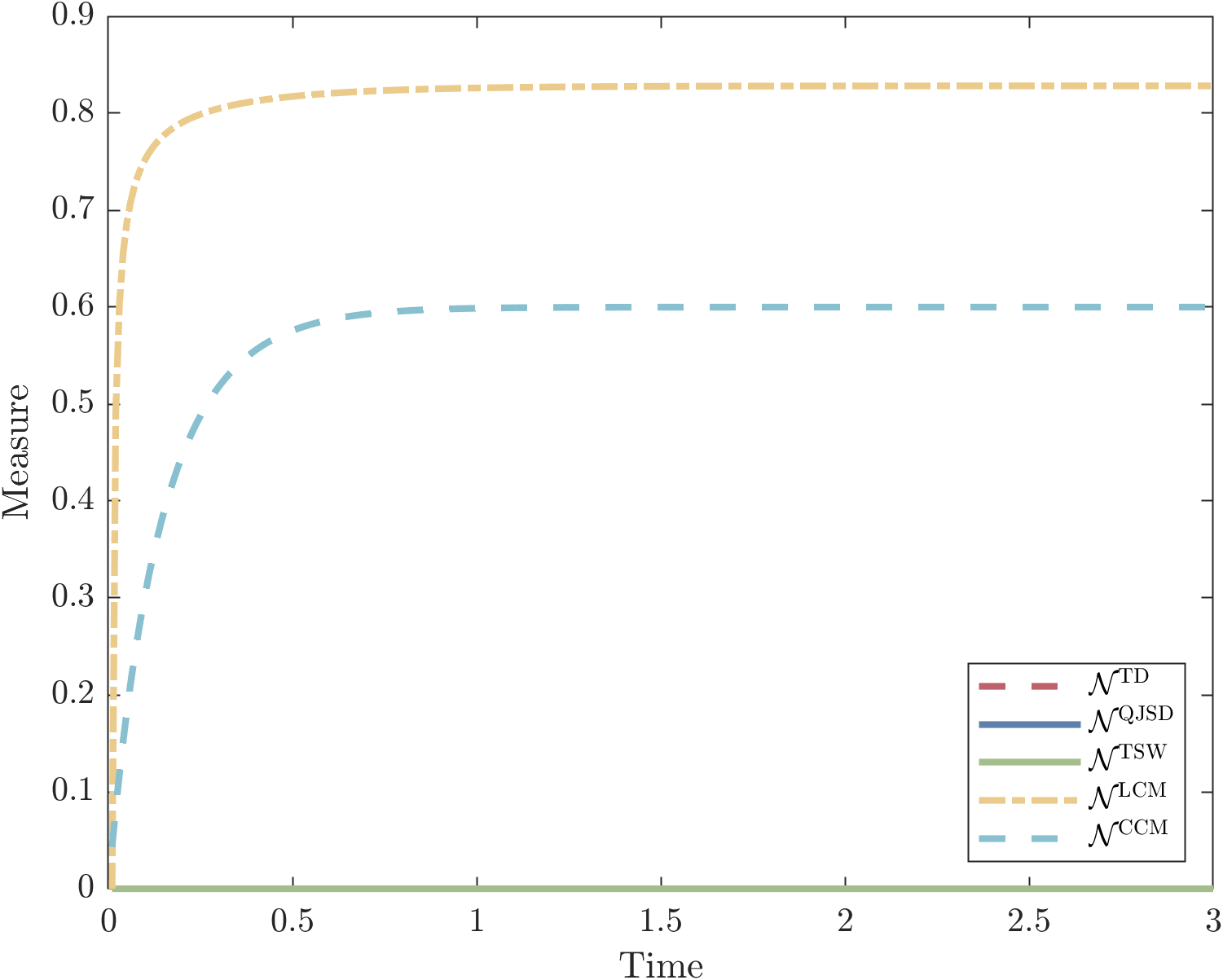}
\caption{The measures of Non-Markovianity for the Eternally non-Markovian channel as described in Eqs.~\eqref{eq:ENM-channel} and \eqref{eq:ENM}. This plot uses $c = 3.0$. It is seen that this channel is considered Markovian by trace distance, temporal steerable weight, and QJSD, but is detected to be non-Markovian by the causality-based measures.}
    \label{fig:ENM-measure}
\end{figure}

It was shown \cite{hall2014canonical} that Choi matrix-based witnesses could detect non-Markovianity of this channel. It is, however, straightforward to see that the CCM-based measure introduced in this work would directly measure the total memory in this channel, but since TSW-based measure is computed numerically, we show through numerical simulation that LCM and CCM-based measure quantify eternal non-Markovianity of this channel, while TSW-based measure does not. This highlights an important difference in the hierarchy of temporal quantum correlations that TSW-based measure may not be equivalent to Choi-matrix based measures while defining a witness of non-Markovianity based on divisibility, even though it was shown that breakdown of monotonicity of TSW is related to negativity of decay rate in the time-dependent Lindblad equation. Here, it is obvious that PDM-based measures capture the eternal non-Markovianity of a P-divisible channel since PDM is proportional to the Choi matrix of a given channel.  

\section{Conclusions \label{sec:discon}} 
We compare distance measures, such a trace distance and entropic distance measures, with the causality measure arising from pseudo-density matrix (PDM) and temporal steerable weight (TSW) which quantifies temporal steering as measures of non-Markovianity. We have shown, through examples and numerical simulations, that temporal steerable correlations are faithful witnesses of information backflow, hence of P-indivisibility, where trace distance and entropic distances may fail \cite{smirne2022holevo,liu2013nonunital,makela2015bounds,bylicka2017constructive,megier2021entropic}. It is known in the literature that distance measures are ineffective in capturing non-Markovianity of processes which do not induce information backflow. We have shown that is also holds for temporal steerable correlations (weak form of quantum direct cause). For example, for an eternally non-Markovian channel that is P-divisible but CP-indivisible, TSW-based based measure is shown to be ineffective hence may not be proportional to Choi matrix-based witnesses. Since PDM-based measures (strong form of quantum direct cause) are shown to be proportional to negativity of decay rates under a non-Markovian channel, they faithfully witness and quantify total memory in any indivisible process. Nevertheless, it is important to point out that the measure based on the trace norm of the Helstrom matrix is known to be shown to be strictly stronger than all entropic measures of non-Markovianity \cite{settimo2022entropic}. As far as detecting information backflow is concerned, we find that PDM-based LCM is found to be stronger than than TSW-based measure in the same parameter regime where QJSD is stronger than trace distance. One must also note that PDM-based causality measures have a specific advantage that they do not require any sort of optimization, while others studied here do. 

Our work takes a step further in understanding the relationship between causality and non-Markovianity in open systems.
   
\acknowledgements
SU thanks IIT Madras for the support through the Institute Postdoctoral Fellowship.

\end{document}